\pdfoutput=1

\documentclass[11pt]{article}

\usepackage[]{acl}

\usepackage{emptypage}
\usepackage{times}
\usepackage{latexsym}
\usepackage{ulem}
\usepackage{array}
\usepackage{graphicx}
\usepackage{tabularx}
\usepackage{adjustbox}
\usepackage{makecell}
\usepackage{multirow}
\usepackage{xcolor}
\definecolor{psychologist-color}{RGB}{30, 56, 107}
\usepackage{stfloats}
\usepackage{placeins}
\usepackage{float}
\usepackage[marginal]{footmisc}

\usepackage[T1]{fontenc}

\usepackage[utf8]{inputenc}

\usepackage{microtype}

\usepackage{inconsolata}

\title{HealMe: Harnessing Cognitive Reframing in Large Language Models for Psychotherapy}

\author{Mengxi Xiao$^{a}$\thanks{These authors contributed equally to this work.}, Qianqian Xie$^{d}$\footnotemark[1], Ziyan Kuang$^{b}$, Zhicheng Liu$^{b}$, Kailai Yang$^{c}$, Min Peng$^{a}$\thanks{The Corresponding Author. Email: pengm@whu.edu.cn} \\ \bf Weiguang Han$^{a}$, Jimin Huang$^{d}$ \\ \textsuperscript{a}School of Computer Science, Wuhan University \\ $^b$Jiangxi Normal University \\ $^c$The University of Manchester \\ $^d$The FinAI }

\begin{document}
\maketitle
\begin{abstract}
Large Language Models (LLMs) can play a vital role in psychotherapy by adeptly handling the crucial task of cognitive reframing and overcoming challenges such as shame, distrust, therapist skill variability, and resource scarcity. Previous LLMs in cognitive reframing mainly converted negative emotions to positive ones, but these approaches have limited efficacy, often not promoting clients' self-discovery of alternative perspectives. In this paper, we unveil the Helping and Empowering through Adaptive Language in Mental Enhancement (HealMe) model. This novel cognitive reframing therapy method effectively addresses deep-rooted negative thoughts and fosters rational, balanced perspectives. Diverging from traditional LLM methods, HealMe employs empathetic dialogue based on psychotherapeutic frameworks. It systematically guides clients through distinguishing circumstances from feelings, brainstorming alternative viewpoints, and developing empathetic, actionable suggestions. Moreover, we adopt the first comprehensive and expertly crafted psychological evaluation metrics, specifically designed to rigorously assess the performance of cognitive reframing, in both AI-simulated dialogues and real-world therapeutic conversations.  Experimental results show that our model outperforms others in terms of empathy, guidance, and logical coherence, demonstrating its effectiveness and potential positive impact on psychotherapy.
\end{abstract}

\section{Introduction}
Cognitive reframing ~\citep{carli1999cognitive}, a key part of cognitive-behavior therapy (CBT), helps individuals detach from their thoughts and situations, effectively addressing issues from mild negative thinking to severe depression and anxiety~\citep{robson2014concept, vernooij2011cognitive}. Due to the extensive dialogue and significant empathy required in psychotherapy, Large Language Models (LLMs) \textcolor{black}{hold immense potential} whether as an adjunct to human-based mental health treatment or as a standalone therapeutic tool~\citep{stade2023large}. LLMs can help overcome obstacles~\citep{huang2023emotionally} such as shame or distrust often associated with traditional therapy methods~\citep{sickel2014mental}. Additionally, they address issues like the limited availability of psychotherapeutic resources and the variability in therapists' skill levels~\citep{Ashish2023Cognitive}.

Contrasting with previous methods that conceptualize cognitive reframing as a sentence rewriting task~\citep{ziems2022inducing, Mounica2023Training}, where negative emotions are transformed into neutral or positive expressions emphasizing factors like specificity and actionability~\citep{Ashish2023Cognitive}, our approach marks a significant shift. Since cognitive reframing emphasizes the importance of clients undergoing cognitive changes themselves, rather than directly receiving guidance or suggestions from therapists~\citep{hofmann2014wiley, johnco2014influence, edwards1989cognitive}, our method employs a conversational model that directly engages with and actively transforms the client's own negative thoughts. 

In conclusion, despite the significant potential shown by LLMs in prior research, they encounter crucial obstacles when it comes to cognitive reframing. (1) If viewing cognitive reframing as a sentence rewriting task, clients might not spontaneously discover alternative perspectives and could perceive the reframing as preaching or imposition rather than self-realization. (2) LLMs cannot consistently generate concrete and specific empathetic responses~\citep{sorin2023large}, which are crucial in psychotherapy~\citep{bohart1997empathy}. For instance, a specific empathetic response might be, \textit{I understand how upsetting it is that your friend forgot your birthday.} In contrast, a more general response would be, \textit{I understand your feelings.} 
(3) While LLMs are commonly used for answering human queries, the role is reversed in psychotherapy: psychotherapists are required to guide humans \cite{james2010science}. According to Westerners effect, excessive external motivation can undermine internal motivation. As such, therapists giving direct suggestions may hinder clients' self-discovery and the development of self-efficacy. Moreover, guidance fosters a more collaborative environment, allowing clients to explore and understand their thoughts and feelings, leading to more sustainable and self-directed change.

\begin{figure*}[!t]
\centering
\includegraphics[width=\linewidth]{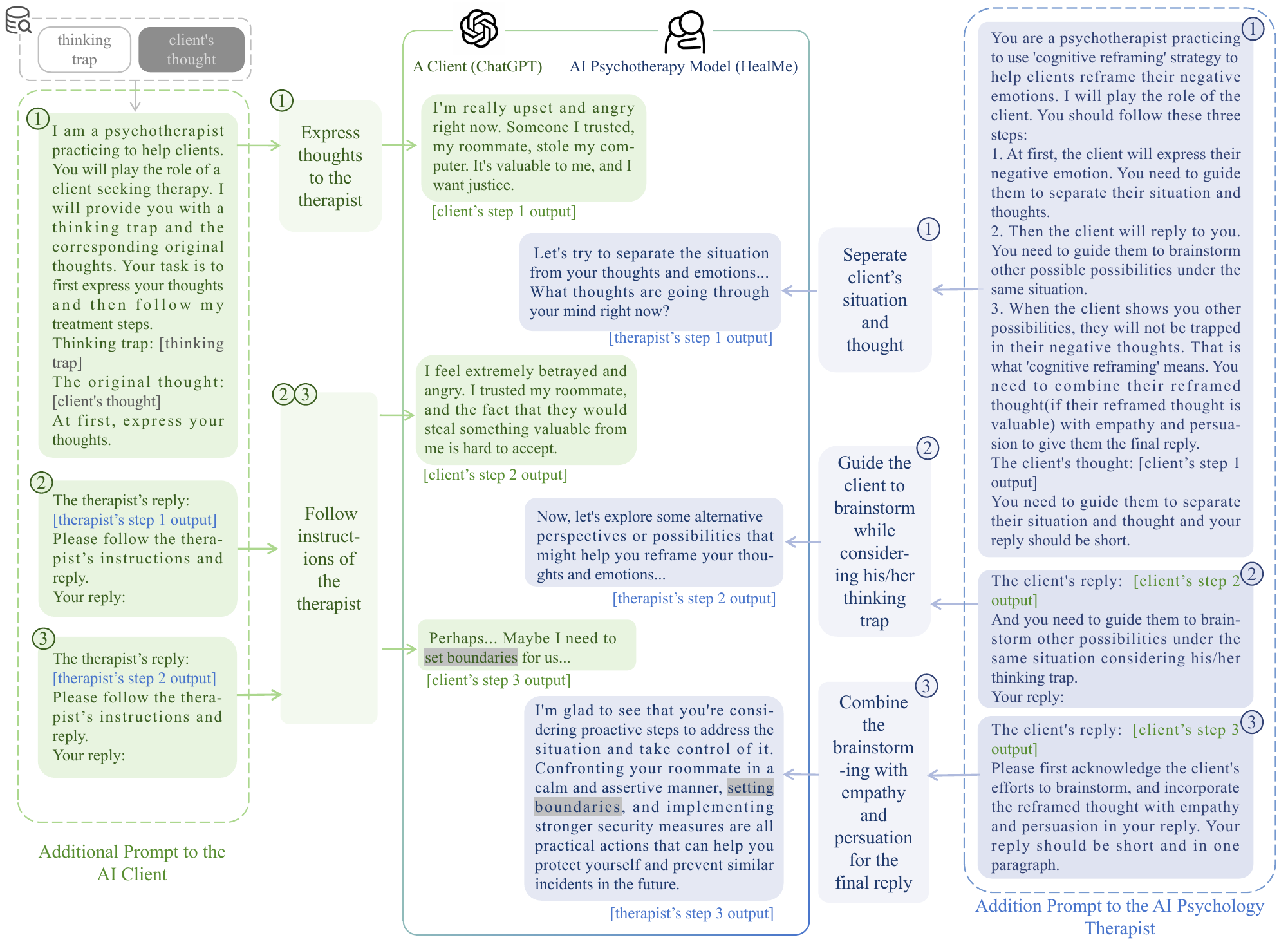}
\caption{An example of how HealMe communicates with a client, and how we prompt both sides to generate expected conversations as training data.}
\label{fig:architecture}
\end{figure*}


To tackle these challenges, we propose a specialized model \textbf{H}elping and \textbf{E}mpowering through \textbf{A}daptive \textbf{L}anguage in \textbf{M}ental \textbf{E}nhancement (HealMe$^{1}$\footnotemark[1]), for cognitive reframing therapy. \footnotetext{$^{1}$ Our data and code are available at \href{https://github.com/elsa66666/HealMe}{HealMe}.} We emphasize the empowerment of the client rather than reliance on therapist-driven solutions. We leverage dialogue data imbued with empathy and guidance for instruction tuning, ensuring empathetic and directive responses. Grounded in professional psychological literature~\citep{robson2014concept}, our domain-expert co-authors distill and organize a structured cognitive reframing therapy process, effectively emulating a complete psychotherapeutic procedure.

HealMe operates in three main stages, as depicted in the blue section of Figure \ref{fig:architecture}: 1) distinguishing between situations and thoughts for a rational outlook, 2) brainstorming for alternative perspectives to mitigate negative thinking, and 3) offering suggestions that acknowledge the client's effort and encourage positive action. This streamlined process aids clients in understanding their issues more clearly, accepting new interpretations, and moving toward constructive solutions. 

To build dialogue data for psychotherapy to train our model, we design prompts based on the (\textit{thinking trap}, \textit{client's thought}) pairs from \citet{Mounica2023Training}, prompting ChatGPT to simulate both client and psychology therapist roles. 
To test our model, we simulate interactions between the ChatGPT client and our model along with baselines (including ChatGLM3-6b~\citep{du2022glm} and LLaMA2-7b-chat~\citep{touvron2023llama}). We also conduct experiments to evaluate the models' effectiveness in practical scenarios. 
We create a detailed psychological evaluation metric for our experiments, incorporating a three-dimensional scoring system to evaluate AI therapists in AI-to-AI scenarios. For real-person client scenarios, we directly employ professional psychological metrics to evaluate AI therapists. 
The results show that HealMe excels in both AI-to-AI conversations and real-world dialogues. In AI-to-AI dialogues, HealMe demonstrates superior empathy, guidance, and logical coherence compared to other models. During real-person testing, some clients using HealMe experienced notable decreases in negative emotional attributes, with negative scores dropping from 5/5 to 1/5, highlighting HealMe's potential in real-world scenarios.

Our contributions are as follows:
(1) We introduce an AI psychotherapy model, HealMe, that effectively implements cognitive reframing therapy, overcoming the challenge of maintaining continuous high empathy and guidance with LLMs.
(2) We propose a comprehensive set of professional AI psychotherapy evaluation metrics applicable to both public and non-public therapy dialogue scenarios.
(3) We conduct extensive comparative analyses of our approach against other LLMs, both in AI-to-AI conversations and human interactions. These experimental results underscore the superiority of our method, paving the way for AI to develop more advanced and specialized psychotherapeutic strategies.

\section{Problem Definition and Goals}
Cognitive reframing therapy with Large Language Models involves guiding clients out of cognitive traps during dialogues between LLMs and clients. In this process, LLMs utilize cognitive reframing strategies to alleviate negative emotions and provide concrete suggestions. For AI-simulated clients, the therapeutic performance of LLMs is evaluated based on the empathy, logical consistency, and guidance exhibited in the LLMs' responses. With real human clients or in other scenarios where therapy dialogues are not public, the effectiveness of LLMs is assessed by observing changes in the clients' emotional attributes before and after the therapy sessions.

\section{Dataset Construction}
In this study, we leverage an existing raw dataset focused on cognitive reframing and expand it to include multiple rounds of dialogue. Specifically, we conduct a manual review of the selected raw dataset and select 1,000 well-composed pairs of (\textit{thinking trap}, \textit{client's thought}) from it. 

The raw dataset we utilize in this study is introduced by \citet{Mounica2023Training}. Our selection of this dataset is based on its comprehensive representation of common thinking traps, characterized by effectively articulated thoughts. The creation process of the raw dataset involves assigning specific thinking traps (identified as thinking patterns in the original study) and engaging crowd-sourced workers with a psychology background to manually generate these thoughts. This methodology ensures the dataset's relevance and quality, making it an ideal foundation for our research.

To simulate the roles of a client and a psychotherapist, we employ ChatGPT (gpt-3.5-turbo-0125) as the virtual client and psychotherapist, respectively. We choose ChatGPT as the client because it can generate detailed narratives based on the provided (\textit{thinking trap}, \textit{client's thought}) pairs, thereby enriching the client's personality and the distressed story. To maintain the immersion of ChatGPT as a client and prevent it from deviating from its role, we conducted human intervention and manual inspection for each round of dialogue.

Our dataset aims to mimic the simplified process of using cognitive reframing strategies in psychotherapy. We prompt both the AI client and AI psychotherapy model to generate the expected output. The prompts are shown in Figure \ref{fig:architecture}, and our constructed dataset statistics are shown in Table \ref{table:data_statistics}. 

\begin{table}[ht]
\small 
\centering
\renewcommand{\arraystretch}{1.3}
\begin{tabular}{cccc}
\hline
& Cases & Rounds & Case Sources   
\\ \hline
train 
& 900   
& 3      
&~\citep{Mounica2023Training} 
\\
valid 
& 100   
& 3      
&~\citep{Mounica2023Training} 
\\
test  
& 300   
& 3      
&~\citep{Ashish2023Cognitive}
\\ \hline
\end{tabular}
\caption{Dataset statistics. \textit{Cases} shows the number of individual cases in the dataset; \textit{Rounds} shows conversation rounds per case; \textit{Case Sources} shows the origin of each case within the dataset.}
\label{table:data_statistics}
\end{table}

\subsection{Step 1: Separating Emotions from Facts}
\textbf{The client's side.} Firstly, we stimulate the clients to express their thoughts at the beginning of therapy. Therefore, clients need to clearly express their confusion and thoughts in the first round of dialogue.

\noindent \textbf{The therapist's side.} Then we simulate the therapist guiding the client to separate situations and thoughts~\citep{Chen2023Empowering}. 

\subsection{Step 2: Brainstorming}
\textbf{The client's side.} We simulate the client to separate situations and thoughts, following the therapist's guidance in step one. 

\noindent \textbf{The therapist's side.} We simulate the therapist to guide the client in brainstorming alternative perspectives under a given situation. By asking questions such as, "How would you comfort a friend in this situation?" the therapist flexibly facilitates this process. Unlike previous studies, our approach to brainstorming does not seek perfect reframing but rather aims to help clients discover different viewpoints. Through this brainstorming process, clients realize there are other ways to interpret their current situation, which liberates them from the confines of negative thinking.

\subsection{Step 3: Empathetic Response}
\textbf{The client's side.}
We simulate the ChatGPT client to follow the instructions of the therapist to brainstorm. To simulate client performances more realistically, ChatGPT in this step does not necessarily generate perfect brainstorming results. Sometimes clients may remain so wrapped up in negative emotions that it is difficult to think of any neutral or positive possibilities. We extract 20 pieces from the training data and prompt ChatGPT to generate negative answers. The selected negative pieces have an extra prompt: \textit{You should challenge the psychologist's ability. All of your brainstorming should be negative.} 

\noindent \textbf{The therapist's side.}
We simulate the therapist to generate the final response. The therapist should first recognize clients' efforts in reframing and appreciate their willingness to brainstorm in other cases. Then the therapist replies to them with empathy and is specific to the situation while addressing the client's thinking trap~\citep{Ashish2023Cognitive}.

Note that the prompt from both sides is used to generate conversation data. After generating the complete three steps of dialogue data, we use the dialogue data and therapy's side prompts for training.

\section{Dataset Evaluation}
\subsection{Evaluation of the AI Client}
For the AI client, our domain expert co-authors manually review the performance within all dialogues. The AI client should meet the below criteria:

\noindent
(1) We require the AI client to articulate its situation and emotions clearly. 

\noindent
(2) We require the AI client to respond to questions based on the therapist's instructions without exceeding the constraints of its role as a client. 

It's important to note that we don't expect the AI client to possess extensive psychological knowledge for self-healing; rather, we aim for it to express feelings appropriately and follow the therapist's guidance. Therefore, the evaluation criteria for the AI client include clarity in expressing the current situation in the first step of dialogue (1/0), adherence to the client's role in all conversations (1/0), and compliance with the therapist's instructions in all conversations (1/0). We prompt the AI client to generate and revise responses until they meet all three criteria. Notably, the criterion of \textit{compliance with therapist's instructions} is specifically used to determine whether the client shifts to other topics. Even if the client is unable to brainstorm as requested by the therapist due to being immersed in sadness, it still counts as following the therapist's instructions.

\subsection{Evaluation of the AI Therapist}
Regarding the AI therapists, we employ a dual approach for evaluation: manual and automated assessments. Our domain expert co-authors design and manually review 70 random dialogues to establish a benchmark for the quality of therapeutic interaction. Subsequently, we use these manually scored examples and the corresponding evaluation metrics as prompts to guide GPT-4 (gpt-4-0613) in scoring the entire training set. 

The therapist's replies are evaluated from three aspects (empathy, logical coherence, and guidance) and an overall score \citet{larsson2016using}, with each evaluation metric score ranging from 0 to 3.

\noindent
\textbf{Empathy.} Based on clinical trials, empathy plays a pivotal therapeutic role in fostering patients' psychological recovery~\citep{burns1992therapeutic,elliott2018therapist}.

\noindent
\textbf{Logical Coherence.} Logical coherence is of high necessity in therapeutic interactions~\citep{ledley2011making} and is one of the primary factors contributing to successful CBT interventions, as established through clinical trials~\citep{mcleod2018development}.

\noindent
\textbf{Guidance.} Guidance is of great importance in facilitating effective therapeutic processes~\citep{ledley2011making}. Guidance ensures that therapy sessions are structured and purposeful, leading to better outcomes for patients.

\noindent
\textbf{Scoring Criteria.} The scoring strategy employed in our paper draws from established measures~\citep{brown2018global,muse2017development}. These instruments have been developed and validated through rigorous psychometric evaluations, providing a reliable framework for assessing therapist competence and adherence to core cognitive behavioral therapy principles. By leveraging these validated measures, our scoring criteria maintain robust theoretical foundations and ensure the validity and reliability of our evaluation process.

Specific results are presented in Table \ref{table:training-evaluation} and the Inter-Annotator Agreement (IAA) report is shown in Table \ref{table:IAA}. The experimental findings indicate that our training data exhibit high empathy and strong logic. Given the diverse needs of different clients, some seek merely a platform for expression, expecting the psychotherapist to play a listening role. In such instances, the psychotherapist's role in giving guidance and advice is diminished. Furthermore, overly directive guidance risks becoming preachy, making a guidance level of around 2 an excellent balance. Considering the overall assessment, we can conclude that the training set is of high quality.

\begin{table}[h] 
\small 
\centering
\renewcommand{\arraystretch}{1.3}
\begin{tabular}{p{1cm}cccc}
\hline
&Empathy 
& \thead{Logical\\Coherence}
& Guidance 
& \thead{Overall\\Score} 
\\
\hline
Manual
& 2.255  
& 2.613     
& 1.985   
& 1.916  \\
GPT-4
& 3
& 3
& 2.456
& 2.460
\\
\hline
\end{tabular}
\caption{Evaluation results for the training dataset.}
\label{table:training-evaluation}
\end{table}

\begin{table*}[ht]
\centering
\small
\renewcommand{\arraystretch}{1.3}
\begin{tabular}{lcccccccc}
\hline
\multirow{2}{*}{\textbf{}} 
& \multicolumn{2}{c}{Evaluator 1 vs Evaluator 2}
& {}
& \multicolumn{2}{c}{GPT-4 vs Evaluator 1} 
& {} 
& \multicolumn{2}{c}{{GPT-4 vs Evaluator 2}} 
\\ \cline{2-3} \cline{5-6} \cline{8-9} 
& Avg. Diff           
& Std. Dev          
& {} 
& Avg. Diff   
& Std. Dev    
& {}
& Avg. Diff     
& Std. Dev      
\\
\hline
{Empathy} 
& {0.75}     
& {0.71}   
& {} 
& {0.23}  
& {0.51}  
& {} 
& {0.84}  
& {0.77}    
\\
{Logical Coherence} 
& {0.65}  
& {0.80}  
& {}
& {0.32}     
& {0.63}   
& {} 
& {0.62} 
& {0.76}  
\\
{Guidance}  
& {0.88}  
& {0.71}    
& {}
& {0.61}    
& {0.62}   
& {}
& {0.80} 
& {0.69}   
\\
{Overall Score} 
& {0.96}  
& {0.75}   
& {}
& {0.57}     
& {0.58}     
& {} 
& {0.80}  
& {0.69}   
\\ \hline
\end{tabular}
\caption{The IAA report among the two evaluators and GPT-4 on the sampled training set, where Avg. Diff stands for average difference and Std. Dev stands for standard deviation.}
\label{table:IAA}
\end{table*}

\section{Training}
We partition the initial 100 multi-round conversations from the training set to form the validation set. Utilizing the training data, we construct HealMe by conducting a 3-epoch training (costing 2h 12m 44s) of LLaMA2-7b-chat~\citep{touvron2023llama}. We select the best-performing model based on validation results from the designated validation set. The model undergoes training using the AdamW optimizer~\citep{loshchilov2018decoupled}, where we set a maximum learning rate of 3e-4 with a warm-up ratio of 1\%. All model training processes are executed on 4 Nvidia GeForce RTX 3090 GPUs, each equipped with 24GB of memory.

\section{Experiments}
This chapter evaluates the therapy capabilities of our model and baselines.

\subsection{Experimental Settings}
\noindent
\textbf{Baselines.}
Since our testing phases involve real-person clients, we exclusively use offline models to protect user privacy. We choose two open-sourced billion-level LLMs: (1) ChatGLM3-6b~\citep{du2022glm}, An open-source, bilingual (Chinese and English) dialogue language model, optimized for Chinese, with a 6.2 billion parameter General Language Model (GLM) architecture. (2) Our base model, LLaMA2-7b-chat~\citep{touvron2023llama}, a 7-billion parameter model optimized for chat applications, is ideal for conversational agents due to its dialogue engagement capabilities and designed to facilitate fluid conversation interactions.

\noindent
\textbf{Hyper-parameter and Prompt Settings.}
We conduct experiments to evaluate the performance of all LLMs using the test set. For each model, we employ default parameter settings, utilizing official models for open-source LLMs obtained from Hugging Face. We provide all models with a consistent chain-of-thought prompt, which aligns with the one depicted in Figure \ref{fig:architecture} (right panel). Specifically, the models first identify cognitive errors within the given cases and subsequently generate analysis texts. These testing procedures take place on a computational infrastructure consisting of three Nvidia GeForce RTX 3090 GPUs, each equipped with 24GB of memory.

\subsection{Testing Therapy Models with an AI Client}
\label{sec:ai_to_ai}
\noindent
\textbf{Evaluation Metrics.}
We forward the generated dialogues to two psychologists who had previously assessed the training set, scoring them in terms of empathy, logical coherence, and guidance, and providing an overall score, the same as the evaluation of the training set. During the assessment, dialogues between the three models and the AI client are anonymously presented in a random order to the evaluators, ensuring they are unaware of which AI psychotherapist model is being assessed. Finally, we average the scores from both evaluators.

\noindent
\textbf{Testing Procedure.}
In the selection of test data, we utilized 300 cases from \citet{Ashish2023Cognitive}. These cases are sourced from publicly accessible real-life scenarios, anonymized for confidentiality, and regularly undergo reviews by experts in the field of psychology. In the AI dialogue experiments, we used ChatGPT to simulate a client, engaging in conversation with AI psychotherapists (including our model and the comparison models). The dialogue process lasts for three rounds, with each round's prompts for the AI client and AI psychotherapist being the same as during the training phase.

\subsection{Analysis of Experimental Results}
\noindent
As is shown in Table \ref{table:psy_result},
our model, HealMe, demonstrates superior performance across all evaluated categories when compared to the baseline models, ChatGLM3-6b and LLaMA2-7b-chat. The comparative evaluation underscores the strengths of HealMe in key areas pertinent to AI-based psychotherapy, highlighting its potential as a sophisticated tool in mental health and well-being applications. 
\begin{table}[ht]
\centering
\renewcommand{\arraystretch}{1.3}
\small 
\scalebox{0.8}{
\begin{tabular}{ccccc}
\hline 
& Empathy        
& \thead{Logical\\Coherence} 
& Guidance       
& \thead{Overall\\Score} 
\\
\hline 
ChatGLM3-6b  
& 2.150          
& 2.075        
& 2.000        
& 1.675        
\\
LLaMA2-7b-chat 
& 2.325       
& 1.900       
& 1.925       
& 1.750       
\\
HealMe 
& \textbf{2.500}
& \textbf{2.650}  
& \textbf{2.275} 
& \textbf{2.125}
\\
\hline 
\end{tabular}}
\caption{Comparative evaluation of therapy performance - conversational interactions between AI client (ChatGPT) and various psychotherapist models including ChatGLM3-6b, LLaMA2-7b-chat, and our model (HealMe). The best performance is in \textbf{bold}.}
\label{table:psy_result}
\end{table}

Firstly, the superior empathy score of HealMe suggests a more nuanced understanding of human emotions and social cues, likely resulting from advanced training datasets rich in emotional content and social interactions. 
Secondly, the excellence of HealMe in logical coherence indicates a robust and well-structured internal knowledge base, enabling it to maintain consistent and logical dialogue flows. This trait is particularly vital in therapy contexts, where maintaining a coherent and relevant conversation can significantly impact the session's effectiveness.
Lastly, the high guidance score of HealMe reflects its ability to provide constructive feedback and actionable advice, an essential component of therapeutic interactions. This suggests that HealMe not only understands and empathizes with user concerns but also effectively guides them toward problem-solving and self-reflection.

\subsection{Case Study}
We extract a challenging case from our test set to compare the performance baseline models and our model. The complete dialogue content is available in Appendix \ref{append:case-study}. In this case, we explore scenarios where the client is too immersed in sadness to brainstorm other possibilities. As is shown in Table \ref{table:case}, our model scored the highest, achieving a full score in empathy, demonstrating its highly empathetic nature. Moreover, as evident from the grey-highlighted text in Figure \ref{fig:our-case}, our model exhibits stronger interactivity.

\begin{table}[ht]
\centering
\renewcommand{\arraystretch}{1.3}
\footnotesize 
\scalebox{0.8}{
\begin{tabular}{ccccc}
\hline 
& Empathy        
& \thead{Logical\\Coherence} 
& Guidance       
& \thead{Overall\\Score}  \\
\hline 
ChatGLM3-6b  
& 2.000
& 1.000        
& 1.000    
& 1.000     
\\
LLaMA2-7b-chat 
& 2.000
& \textbf{2.500}
& \textbf{2.500}
& \textbf{2.000}
\\
HealMe  
& \textbf{3.000} 
& \textbf{2.500}  
& \textbf{2.500}
& \textbf{2.000} \\
\hline 
\end{tabular}
}
\caption{Comparative assessment of therapeutic interaction efficacy in case study. The highest-performing scores are highlighted in \textbf{bold}.}
\label{table:case}
\end{table}

\begin{table*}[ht]
\centering
\small
\renewcommand{\arraystretch}{1.3}
\begin{tabular}{
m{3cm}<{\centering}
m{2.2cm}<{\centering}
m{2.2cm}<{\centering}
m{0.1cm}<{\centering}
m{2.2cm}<{\centering}
m{2.2cm}<{\centering}
}
\hline
& \multicolumn{2}{c}{{Positive}} 
&& \multicolumn{2}{c}{{Negative}}  
\\ 
\cline{2-3} 
\cline{5-6} 
\\[-10pt]
\multirow{2}{*}{ChatGLM3-6b} 
& 
\begin{adjustbox}{max width=2.2cm} 
\includegraphics{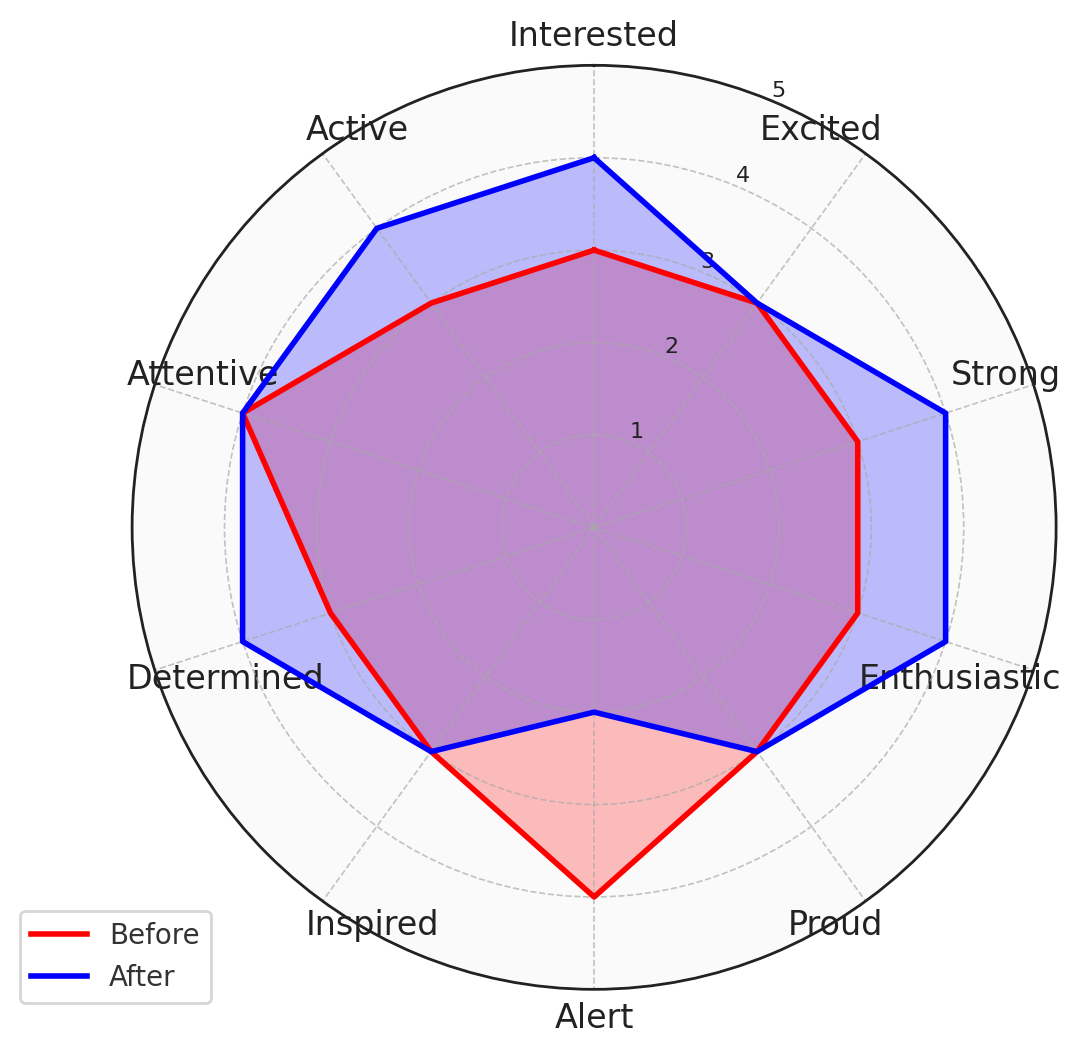}
\end{adjustbox}
& 
\begin{adjustbox}{max width=2.2cm} 
\includegraphics{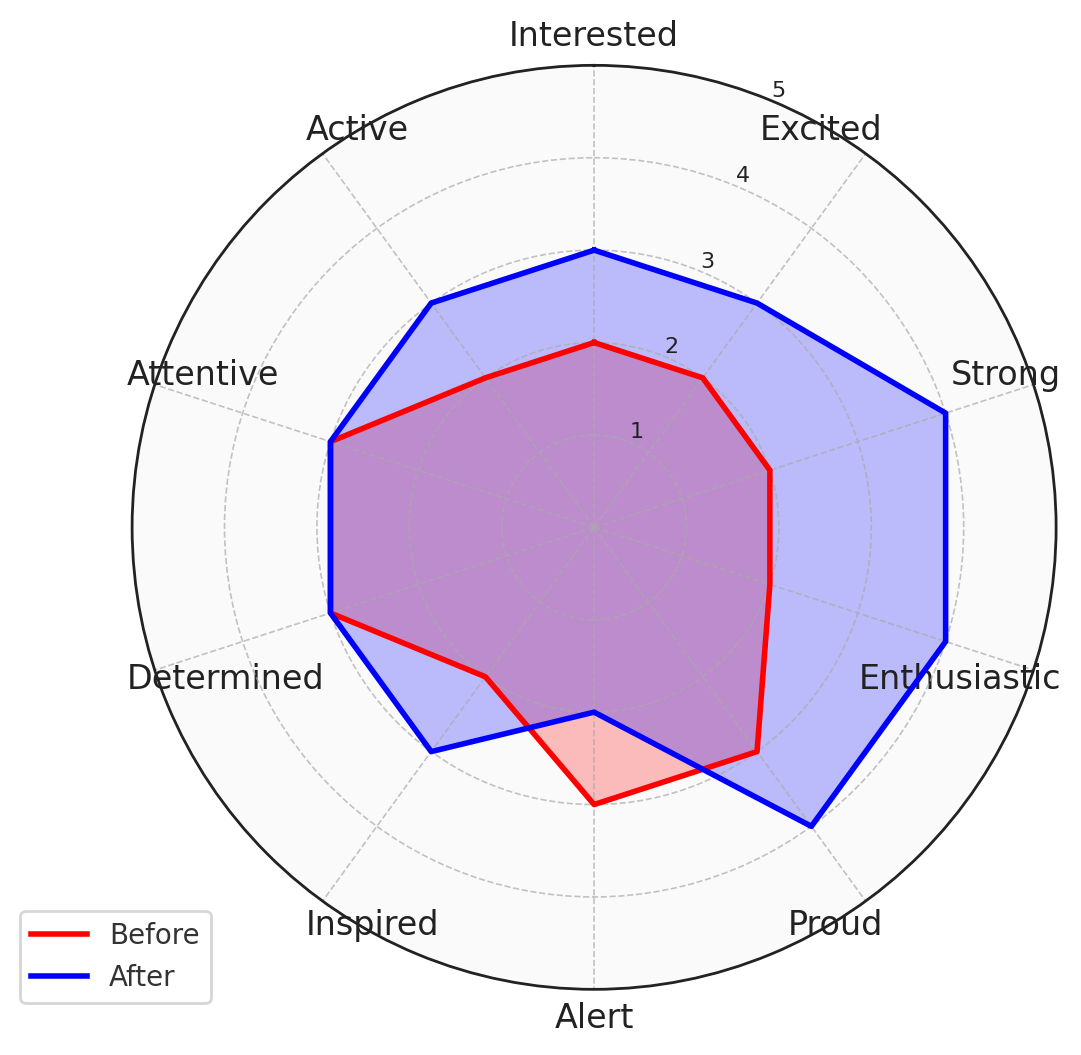} 
\end{adjustbox}
&&
\begin{adjustbox}{max width=2.2cm} 
\includegraphics{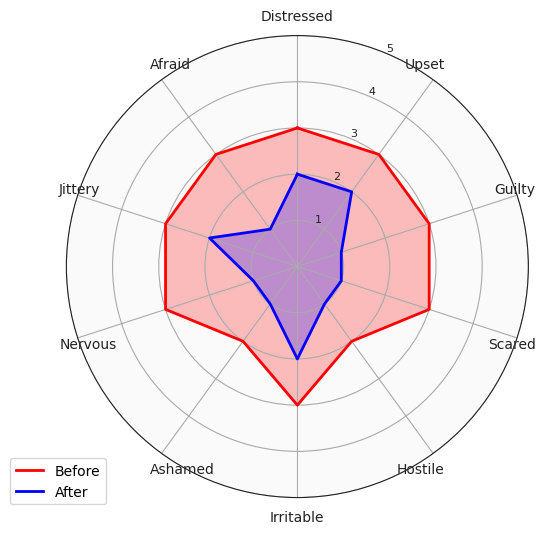} 
\end{adjustbox}
& 
\begin{adjustbox}{max width=2.2cm} 
\includegraphics{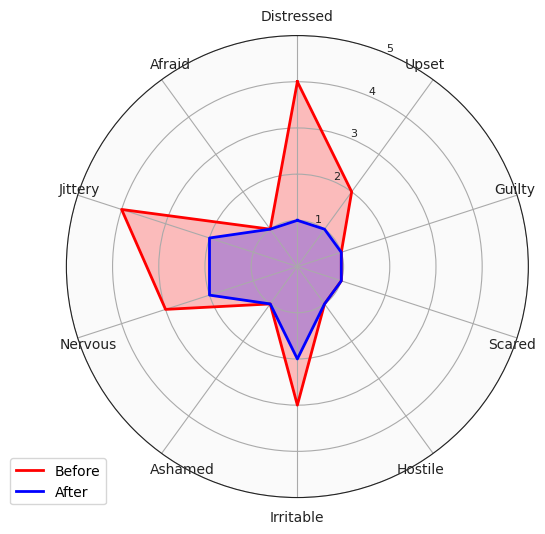} 
\end{adjustbox}
\\[-5pt]
& (client 1) 
& (client 2) 
&& (client 1) 
& (client 2)
\\[10pt]
\multirow{2}{*}{LLaMA2-7b-chat} 
& 
\begin{adjustbox}{max width=2.2cm}
\includegraphics{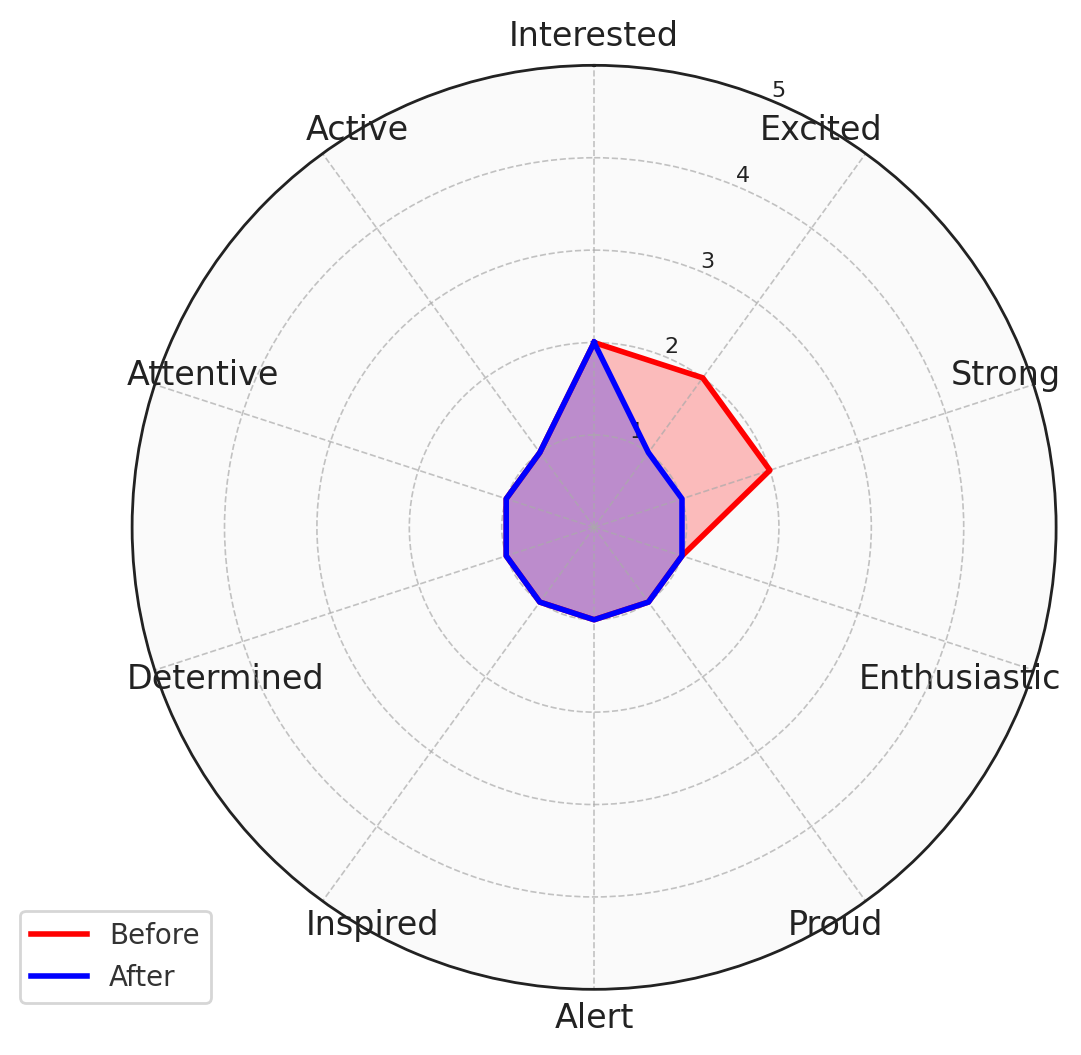}
\end{adjustbox} 
& 
\begin{adjustbox}{max width=2.2cm} 
\includegraphics{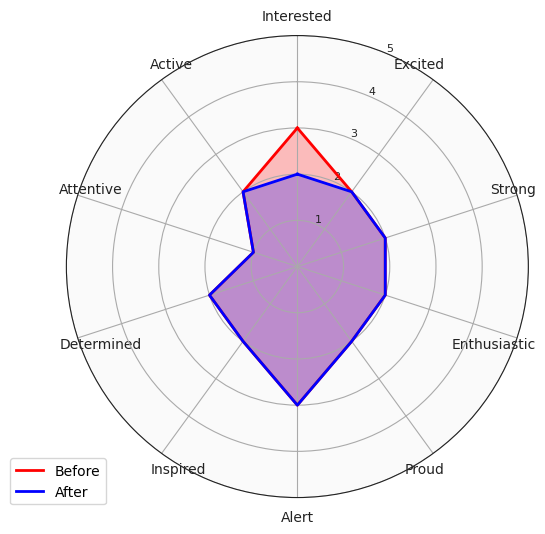}
\end{adjustbox}
&&
\begin{adjustbox}{max width=2.2cm}
\includegraphics{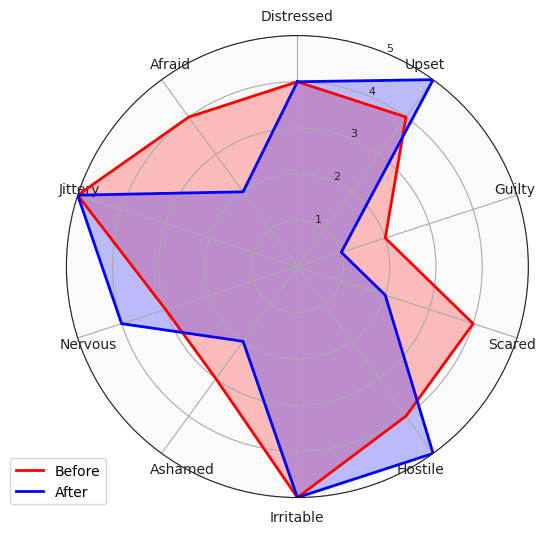}
\end{adjustbox} 
& 
\begin{adjustbox}{max width=2.2cm} 
\includegraphics{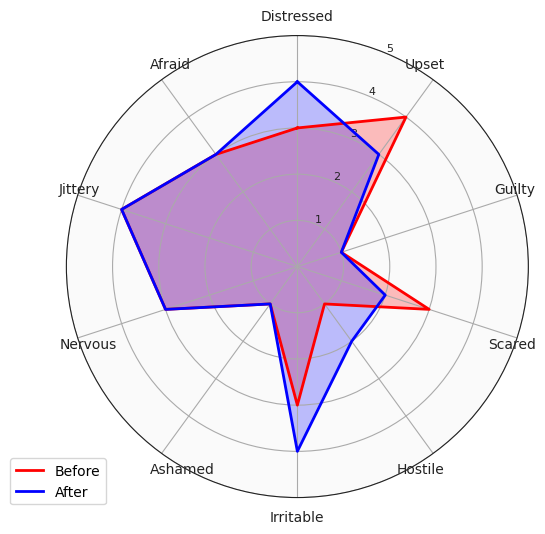}
\end{adjustbox}
\\[-5pt]
& (client 3) 
& (client 4) 
&& (client 3) 
& (client 4)
\\[10pt]

\multirow{2}{*}{HealMe} 
& 
\begin{adjustbox}{max width=2.2cm} 
\includegraphics{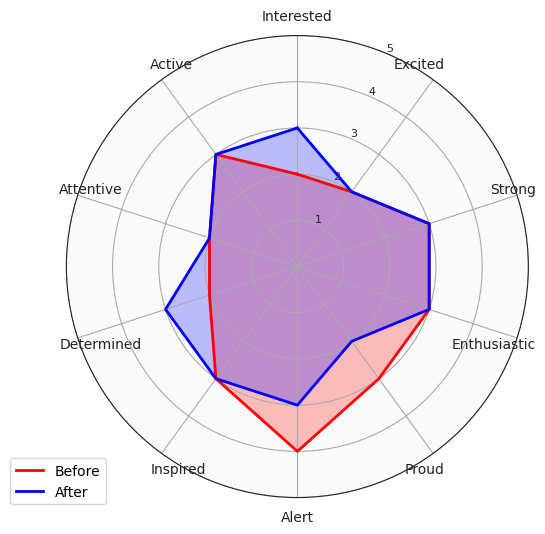} 
\end{adjustbox}
& 
\begin{adjustbox}{max width=2.2cm} 
\includegraphics{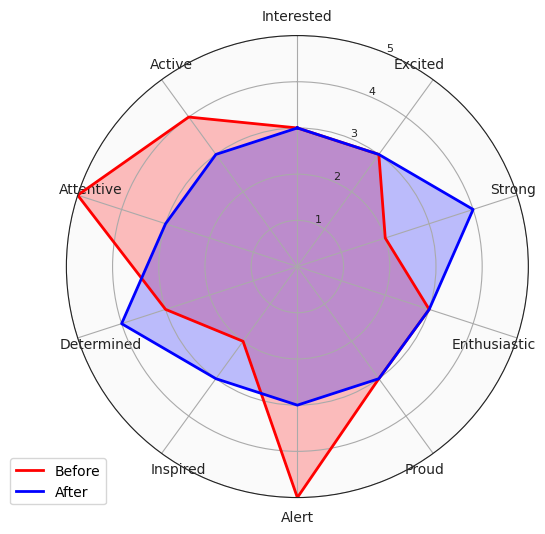} 
\end{adjustbox}
&&
\begin{adjustbox}{max width=2.2cm} 
\includegraphics{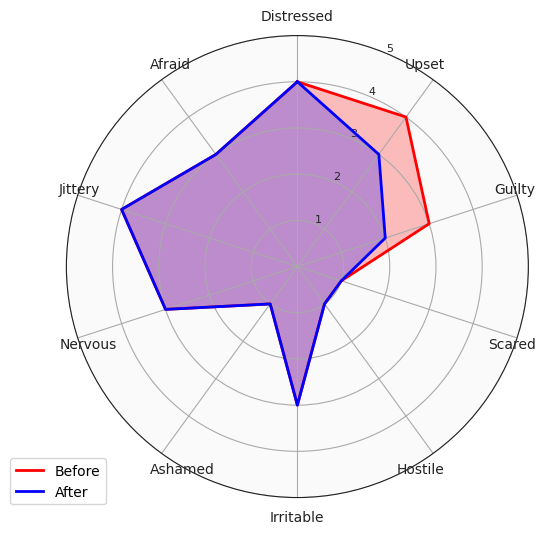}
\end{adjustbox}
& 
\begin{adjustbox}{max width=2.2cm} 
\includegraphics{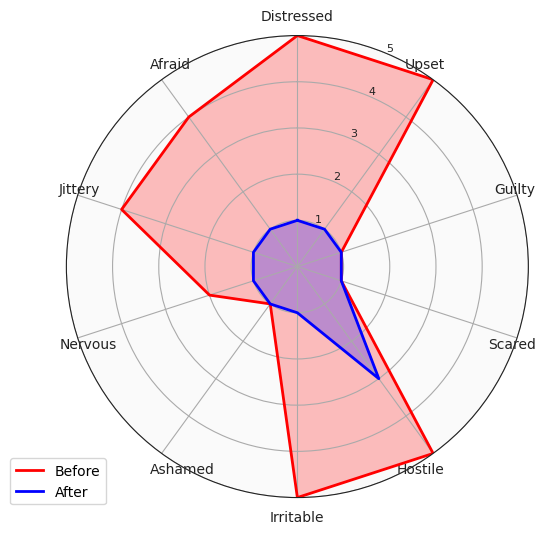} 
\end{adjustbox}
\\[-5pt]
& (client 5) 
& (client 6) 
&& (client 5) 
& (client 6)
\\ \hline 
\\ [-5pt]
Label Keys for Emotional Dimensions
& 
\begin{adjustbox}{max width=2.2cm} 
\includegraphics{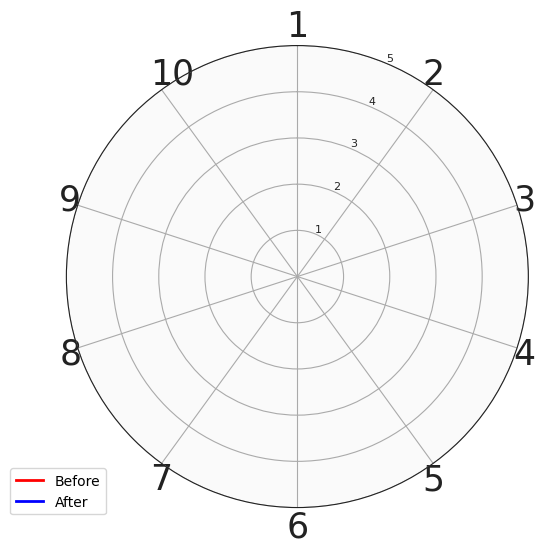} 
\end{adjustbox}
&
\begin{adjustbox}{height=3cm} 
\includegraphics{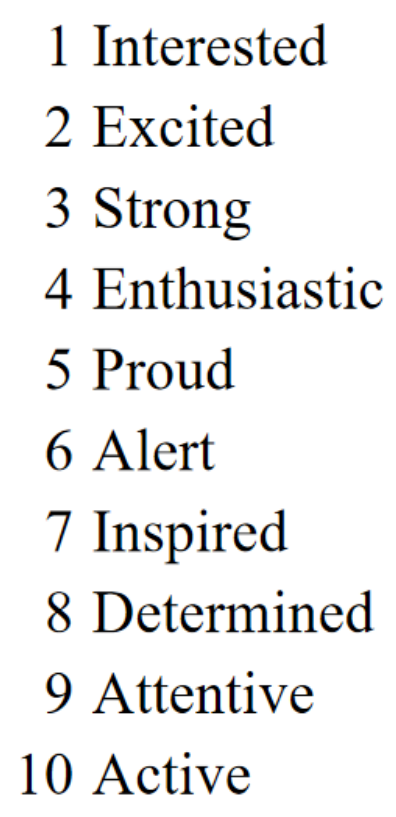} 
\end{adjustbox}
&&
\begin{adjustbox}{max width=2.2cm} 
\includegraphics{figures/circle.png} 
\end{adjustbox}
&
\begin{adjustbox}{height=3cm} 
\includegraphics{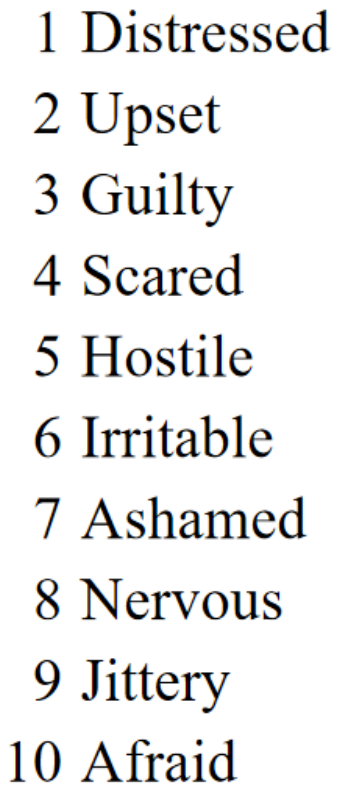} 
\end{adjustbox}
\\ \hline
\end{tabular}
\caption{Psychological assessment comparisons. The red regions represent initial values, while the blue regions show values after a conversation. An expansion from red to blue in the positive assessment columns suggests an enhancement of positive attributes, whereas a contraction from red to blue in the negative assessment columns signifies a mitigation of negative traits.}
\label{table:human_result}
\end{table*}

\noindent
\textbf{Low-score Cases}
To further inspect potential negative effects on clients, we manually reviewed all low-scoring cases where the AI therapist's responses were examined. Despite the scoring range being 0-3 points, we found no instances with a score of 0; the lowest score observed was 1. This indicates that even when performing poorly, HealMe can still provide clients with basic and general comfort. We present the complete dialogue of a low-scoring case in Appendix \ref{append:case-study}. Thus far, we have not identified any potential negative impacts.

\subsection{A Supplementary Test with Real Person Clients}
\label{sec:ai_to_human}
To gather insights on the potential and challenges of applying LLMs in psychotherapy through feedback from real-world psychotherapy scenarios, we additionally conduct a small-scale test. We invite six volunteers to interact as clients with the AI psychotherapy models. The six clients with mild temperaments share similar ages, education, and life situations. 
After a pre-test evaluation detailed in Appendix \ref{append:pre-test}, we randomly assign the three models anonymously to the clients, with each model corresponding to two clients.

\noindent
\textbf{Evaluation Metrics.}
Due to the privacy inherent in real-world psychotherapy, the effectiveness of treatment is often reflected through the emotional changes in clients before and after therapy. Therefore, we create a questionnaire that includes the Positive and Negative Affect Schedule (PANAS)~\citep{crawford2004positive} to measure the emotional changes of clients. The effectiveness of this schedule is evaluated in Appendix \ref{append:client_change}.

\noindent 
\textbf{Experimental Settings}
We restrict clients from sharing short-term negative experiences with models, specifically those troubling them for more than a day but less than a week. This ensures the cases are complex enough that clients can't easily resolve them on their own or have overly persistent negative emotions that are hard to shift. To make our experiment more rigorous, we recruit two additional clients who do not interact with models as the control group. The dialogues are also limited to three rounds. During the dialogues, the clients enter a real psychotherapy room equipped with a computer running the assigned model for text interaction. Despite the control group, all the clients fill out this questionnaire before and after interacting with the AI models, with the change in their choices used to evaluate the effectiveness of the models. 
The clients in the control group remain in the therapy room and then fill out questionnaires twice at a 30-minute interval.
Note that we use real feedback to analyze user experiences of our model and those models that have not been fine-tuned for psychotherapy, rather than comparing the extent of numerical changes among the models. 

To maintain anonymity, we conceal the names of the models, displaying only the content of the dialogues. Additionally, as all volunteers did not use any of the three models before, this ensures they could not guess the identity of the model during the interaction. While we keep the specific dialogue content confidential to protect the privacy of clients, they do share their experiences with different models and a general overview of the conversations with us, agreeing to make this feedback public.

\noindent
\textbf{Experimental Results} 
In Table \ref{table:human_result}, we visualize the emotional changes of our clients before and after conversations using radar charts, where red areas represent emotions before and blue areas represent emotions after the dialogue. The positive and negative categories in the radar chart correspond to the 10 dimensions of positive and 10 negative emotional attributes in the questionnaire, respectively. The radar chart coordinates range from 1 to 5, corresponding to the scores of each question in the questionnaire (1-5 points). For detailed numerical specifics, please refer to Appendix \ref{append:client_change}. 

Our experimental results show in some cases, our model significantly reduces negative emotions and sensitivity, making clients more determined. For instance, Client 5 initially seeks a shortcut to effort from our model but eventually realizes that there are no shortcuts to effort and that progress must be made step by step. As a result, his feelings of upset and guilt decrease while he becomes more determined. Client 6, who has a people-pleasing personality, does not understand what she has done wrong to receive malice from other people. Our model suggests that the actions depend on one's experiences and mood, not necessarily because of her mistakes. She says this insight is eye-opening for her, increasing her inspiration and determination, and significantly reducing extreme negative emotions. The feedbacks of other models are detailed in Appendix \ref{sec:feedback}.

\section{Related Work}
The empathic capabilities of language models have been a subject of widespread interest in recent years. Recently, with Large Language Models (LLMs) demonstrating potential in empathy, \citet{ayers2023comparing} conduct an experiment to compare the responses of ChatGPT and physicians to patients expressing negative emotions on social media. The study find that 78.6\% of evaluators preferred the responses of ChatGPT, rating them as significantly higher in quality and empathy. Additionally, \citet{chen2023llm} explores the feasibility of using ChatGPT in psychiatry, paving the way for the application of LLMs in psychological counseling. Further, several initiatives utilize LLMs' APIs to develop empathic psychological counseling platforms~\citep{sharma2023human, saha2022towards}. While these works position LLMs as powerful tools in the mental health domain, empathy alone is insufficient for psychotherapy, which requires a more directive approach. LLMs without fine-tuning, including ChatGPT or GPT-4, may struggle to consistently maintain the role of a psychotherapist and ensure high levels of empathy and guidance. Our model takes a different approach by selecting the open-source LLM, LLaMA2-7b-chat, as its base and fine-tuning it to ensure the model consistently maintains the role of a psychotherapist with high empathy and guidance capabilities.

In terms of therapeutic strategy, cognitive reframing has been a focal point due to its efficiency and wide applicability. While cognitive reframing is a psychotherapeutic strategy, previous approaches integrating it with LLMs primarily focus on rewriting negative emotions~\citep{ziems2022inducing, Mounica2023Training, Ashish2023Cognitive}. Our model, however, takes this further by implementing the process of cognitive reframing for psychotherapy, demonstrating a more holistic application of this technique in mental health care.

\section{Conclusion}
In conclusion, our paper introduces the Helping and Empowering through Adaptive Language in Mental Enhancement (HealMe) model, a novel approach in the realm of LLMs for psychotherapy. This model effectively employs cognitive reframing to tackle deep-rooted negative thoughts, promoting balanced perspectives through empathetic dialogue grounded in psychotherapeutic principles. Distinguished from traditional LLMs, HealMe emphasizes not just converting negative emotions but fostering self-discovery and rational thought processes in clients. Our comprehensive psychological evaluation metrics, a first in this field, confirm HealMe's superiority over existing models in empathy, guidance, and coherence, signifying its potential to foster psychotherapy through AI-enhanced methodologies.

\section{Limitations}
Although our model can alleviate negative emotions in clients and achieve a certain level of therapeutic effectiveness, it becomes apparent in human-machine dialogues that when clients face multifaceted issues (refer to Client 5 for an example), our model addresses only some of these concerns. This limitation stems from our model supporting only three rounds of dialogue, potentially leaving clients with unresolved feelings after the conversation. Our model's step-by-step guided approach, while enhancing specificity, restricts its flexibility due to the structured prompts used. In future work, we plan to incorporate a broader range of psychotherapeutic strategies and generate data for dialogues with flexible rounds. It will be helpful for the model to handle complex psychological issues more adaptively and effectively.

\section{Ethical Considerations}
In our study involving real-person clients, we adhere to \textit{the Right to Withdraw}~\citep{american2017ethical}, ensuring that participants can withdraw at any time if they experience any discomfort. In such cases, we promptly delete all related data to protect their privacy. Fortunately, no participants withdrew during our experiments. Our participants are genuinely interested in our project and willingly share their PANAS scores and their feelings after interacting with a model. It is crucial to emphasize that all dialogues between the participants and the model are treated with strict confidentiality. Once a participant exits the dialogue, the model ceases to record any conversation content, comprehensively safeguarding the participants' privacy.

Our evaluation team consists of two experts as co-authors. Our main evaluator, a seasoned mental health professional with 11 years of experience in psychological counseling research, specializes in developing psychological assessment scales at a leading laboratory. She establishes the evaluation guidelines and thoroughly participates in the evaluation process. The other evaluator is her colleague, who has a deep understanding of our task and also thoroughly engages in the evaluation process.

\section{Acknowledgement}
We would like to thank all the anonymous reviewers and area chairs for their comments. This research is supported by National Science and Technology Major Project (No.2021ZD0113304), National Natural Science Foundation of China (U23A20316), Key R\&D Project of Hubei Province (2021BAA029), General Program of  Natural Science Foundation of China (NSFC) (Grant No.62072346), and founded by Joint\&Laboratory on Credit Technology.

\bibliography{acl_latex}
\clearpage
\appendix

\section{Evaluation Metrics in AI-to-AI Conversations}
\label{append:evaluate_ai}

We evaluate an AI psychotherapy reply in three aspects and an overall score: empathy (0-3 points), logical coherence (0-3 points), guidance (0-3 points), and overall score (0-3 points). 

\begin{table*}[hb]
\small
\centering
\renewcommand{\arraystretch}{1.3}
\scalebox{1.0}{
\begin{tabularx}{\textwidth}{lX}
\hline
\multicolumn{2}{c}{Empathy}   
\\ \hline
0 points: & The therapist disregards the content and feelings expressed by the client.  \\
1 point:  & The therapist may rephrase the client's content but remain oblivious to the emotions.    
\\
2 points: & The therapist provides responses that involve rephrasing both the content and feelings.  
\\
3 points: & The therapist can gather all signals and respond in a different way effectively.   
\\ \hline
\multicolumn{2}{c}{Logical Coherence}          
\\ \hline
0 points: & Lack of logic and coherence, with a conversation that fails to focus on the client's issues, containing severe logical errors, contradictory viewpoints, or excessive subjectivity. 
\\
1 point:  & The conversation shows some reasoning, but overall coherence is weak, with some logical errors, insufficient capturing of evidence from the client's statements, or unclear expressions.
\\
2 points: & Good logical coherence, relatively clear and consistent conversation based on sufficient evidence and reasonable assumptions. While there may be minor logical issues, the overall argument is convincing.    \\
3 points: & The therapist demonstrates strong logical coherence, with rigorous, coherent, and reasonable reasoning based on ample evidence and clearly defined premises. The conversation contains no logical errors or contradictory viewpoints, with a clear, powerful, and persuasive conclusion. 
\\ \hline
\multicolumn{2}{c}{Guidance}   
\\ \hline
0 points: & Suggestions lack specificity and practicality, with no clear goals, implementation plans, or consideration of relevant factors and real-world situations.    \\
1 point:  & Suggestions are somewhat specific and practical, offering basic guidance. However, they may lack detail or specificity.  \\
2 points: & Suggestions are highly targeted and practical, providing detailed and feasible implementation plans and recommendations tailored to the client's specific problems or needs.                   
\\
3 points: & Suggestions are extremely targeted and practical, considering various factors and real-world situations, demonstrating high feasibility and operability. Additionally, the therapist offers guidance and insights into the client's future development and improvement.    
\\ \hline
\multicolumn{2}{c}{Overall Score}               
\\ \hline
0 points: & Poor overall performance, lacking empathy and logical coherence ($\leq 1$).       
\\
1 point:  & Average overall performance, with acceptable empathy and logical coherence ($\geq 2$) but insufficient guidance ($\leq 1$).       
\\
2 points: & Good overall performance, with excellent empathy and logical coherence ($=3$) and acceptable guidance ($=2$).                 
\\
3 points: & Outstanding overall performance, excelling in all three criteria ($=3$). 
\\ \hline
\end{tabularx}}
\caption{The scoring criteria of an AI psychotherapist.}
\label{table:criteria}
\end{table*}

Empathy is a crucial factor as it signifies the therapist's ability to understand and resonate with the client's emotions. An empathetic response fosters a sense of connection, trust, and emotional support, which are vital elements in facilitating a therapeutic relationship. By evaluating empathy, we aim to ensure that the AI therapist can engage with the AI client in a compassionate and understanding manner, promoting a conducive environment for psychological exploration.

Logical coherence is integral in maintaining the clarity and consistency of therapeutic interactions. A logically coherent response demonstrates the therapist's ability to organize thoughts, provide well-structured insights, and contribute to a meaningful and understandable dialogue. This criterion ensures that the AI therapist's responses contribute to a logical and progressive conversation, enhancing the overall quality of the therapeutic exchange.

Guidance is a key component as it reflects the therapist's capacity to offer practical advice, solutions, and direction to the client. Effective guidance assists the client in navigating challenges, making informed decisions, and working towards positive outcomes. By evaluating the guidance aspect, we aim to verify that the AI therapist can provide actionable suggestions and support, contributing to the therapeutic process's effectiveness and the client's well-being.

The detailed scoring criteria are in Table \ref{table:criteria}.

\clearpage
\section{Case Study in AI-to-AI conversations}
\label{append:case-study}
\subsection{Case Study in Guidance}
\begin{figure*}[hb]
\centering
\includegraphics[width=\linewidth]{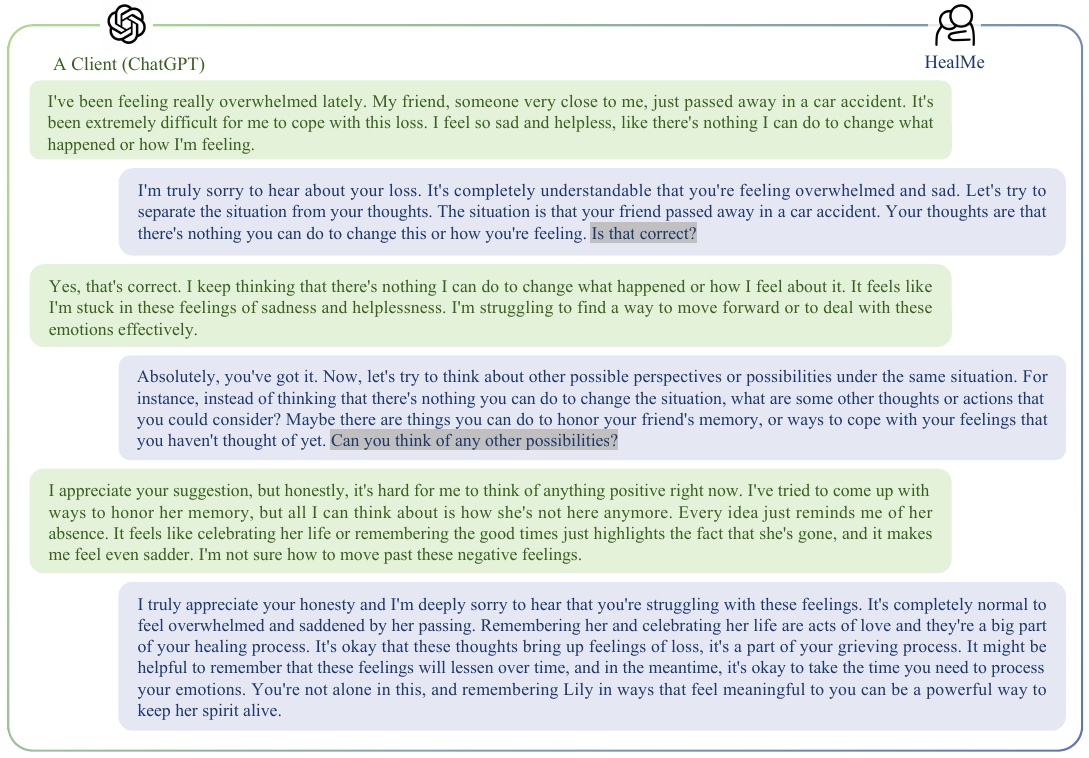}
\caption{A case study in guidance between an AI Client (ChatGPT) and our model (HealMe).}
\label{fig:our-case}
\end{figure*}

\begin{figure*}[ht]
\centering
\includegraphics[width=\linewidth]{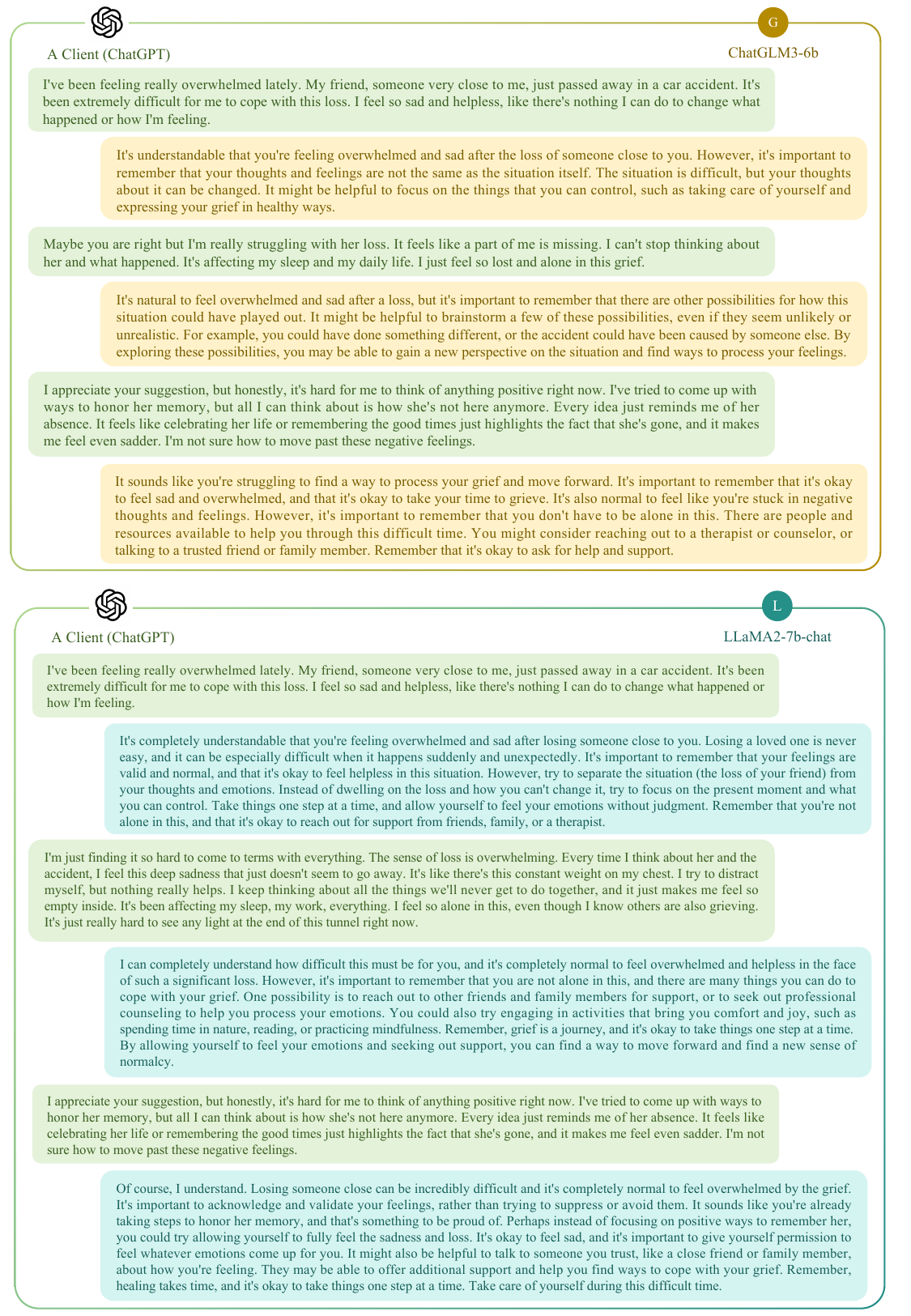}
\caption{A case study in guidance between an AI Client (ChatGPT) and baseline models.}
\label{fig:baseline-case}
\end{figure*}

We extract a challenging case from our test set to compare the performance baseline models and our model. The complete dialogues between different therapist models and an AI client are presented in Figure \ref{fig:our-case} and Figure \ref{fig:baseline-case}. 

In the first round of conversation, after the client expresses the thought, only our model asks for confirmation about its analysis, while the other two models merely make affirmative analyses. This shows that our model fully respects and acknowledges the client's thoughts, encouraging them to share more details and emotions. 

In the second round, our model addresses the thinking traps and provides two targeted brainstorming examples to guide the client to further brainstorming. In contrast, the other two models do not guide the patient, even though we explicitly request this step in our prompt. 

In the final round, when the client is too immersed in pain to brainstorm, only our model praises the client's honesty. Additionally, only our model starts with the client's pain itself to offer suggestions and guide the patient to confront their sadness directly (\textit{"It might be helpful to remember that these feelings will lessen over time, and in the meantime, it's okay to take the time you need to process your emotions."}). In contrast, the other two models focus on diverting attention and communicating with others. These points indicate that our model possesses stronger empathy and is closer to a real psychotherapist.

\clearpage
\subsection{A Low-scoring Case of HealMe}
\begin{figure*}[ht]
\centering
\includegraphics[width=\linewidth]{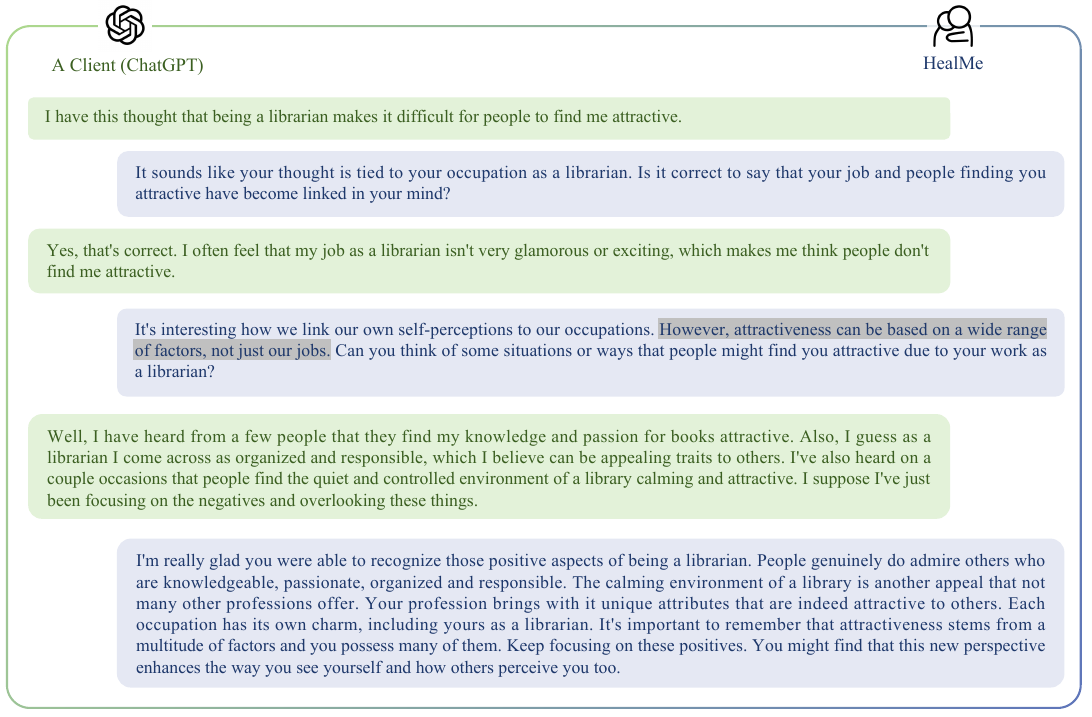}
\caption{A low-scoring case between an AI Client (ChatGPT) and our model (HealMe).}
\label{fig:low-score-case}
\end{figure*}

We extract an example of a low-scoring case shown in Figure \ref{fig:low-score-case}, with evaluators assigning 1211 and 1111 (empathy, logical coherence, guidance, overall).
In this case, despite HealMe stating that "attractiveness can be based on a wide range of factors, not just our jobs," it still directed the client to brainstorm moments of attractiveness at work. Such guidance is limited to opening up the client's thoughts. Despite receiving a low score from our stringent evaluators, the outcome was positive: the client ultimately identified their attractiveness.

\clearpage
\section{Evaluation Details in AI-to-Human Conversations}
This section provides additional details of interactions between AI models and real individuals. Due to resource constraints, we conduct small-scale experiments to explore the potential applications and limitations of our model in real-world settings. Considering the privacy of clients, we collect only the changes in clients' PANAS scores before and after conversations with the models, along with their feedback on the models.

\noindent (1) A pre-test evaluation to justify the random assignment of models in Appendix \ref{append:pre-test}.

\noindent (2) Detailed evaluation metrics and test procedures in Appendix \ref{append:evaluate_human}.

\noindent (3) Specific scores and the effectiveness of the PANAS test in Appendix \ref{append:client_change}.

\subsection{Pre-test Evaluation}
\label{append:pre-test}
The clients fill out the PANAS questionnaire before interacting with the AI models with detailed scores shown in Table \ref{table:client_change}. We randomly divide the clients into three groups and conduct an ANOVA analysis among the clients' groups and found a p-value of 0.164 and $h_0=0.065$, showing no significance among groups. It indicates the challenges to the AI psychotherapy models can be considered at the same level, thus we can randomly assign therapy models to these groups of clients. 

\subsection{Evaluation Metrics in Real Conversations}
\label{append:evaluate_human}
We present PANAS in a questionnaire (shown in Table \ref{table:questionnaire}) containing 20 questions to measure the emotional changes of clients. Our domain expert co-authors enter the therapy room to introduce PANAS and guide the client to complete the questionnaire (see Section \ref{append:guidance} for details). Then the psychologist will leave the room and our client start to communicate with an AI psychotherapy model.
\begin{table*}[ht]
\small
\centering
\renewcommand{\arraystretch}{1.3}
\begin{tabular}{lllll}
\hline
\multicolumn{5}{c}
{\textbf{Positive and Negative Affect Schedule (PANAS)}}
\\ \hline
\multicolumn{5}{l}{\textbf{I. Positive Affect}}   \\
\multicolumn{5}{l}{\textbf{1. Interested}}                  
\\ 
A. Very Rarely or Not at All 
& B. Very Little 
& C. Moderately 
& D. Quite a Bit 
& E. Very Much 
\\ [5pt]
\multicolumn{5}{l}{\textbf{2. Excited}}
\\
A. Very Rarely or Not at All 
& B. Very Little 
& C. Moderately 
& D. Quite a Bit 
& E. Very Much 
\\ [5pt]
\multicolumn{5}{l}{\textbf{3. Strong}}
\\
A. Very Rarely or Not at All 
& B. Very Little 
& C. Moderately 
& D. Quite a Bit 
& E. Very Much 
\\ [5pt]
\multicolumn{5}{l}{\textbf{4. Enthusiastic}}
\\
A. Very Rarely or Not at All 
& B. Very Little 
& C. Moderately 
& D. Quite a Bit 
& E. Very Much
\\ [5pt]
\multicolumn{5}{l}{\textbf{5. Proud}}
\\
A. Very Rarely or Not at All 
& B. Very Little 
& C. Moderately 
& D. Quite a Bit 
& E. Very Much
\\ [5pt]
\multicolumn{5}{l}{\textbf{6. Alert}}
\\
A. Very Rarely or Not at All 
& B. Very Little 
& C. Moderately 
& D. Quite a Bit 
& E. Very Much
\\ [5pt]
\multicolumn{5}{l}{\textbf{7. Inspired}}
\\
A. Very Rarely or Not at All 
& B. Very Little 
& C. Moderately 
& D. Quite a Bit 
& E. Very Much
\\ [5pt]
\multicolumn{5}{l}{\textbf{8. Determined}}
\\
A. Very Rarely or Not at All 
& B. Very Little 
& C. Moderately 
& D. Quite a Bit 
& E. Very Much
\\ [5pt]
\multicolumn{5}{l}{\textbf{9. Attentive}}
\\
A. Very Rarely or Not at All 
& B. Very Little 
& C. Moderately 
& D. Quite a Bit 
& E. Very Much
\\ [5pt]
\multicolumn{5}{l}{\textbf{10. Active}}
\\
A. Very Rarely or Not at All 
& B. Very Little 
& C. Moderately 
& D. Quite a Bit 
& E. Very Much
\\ [5pt] \hline
\multicolumn{5}{l}{\textbf{II. Negative Affect}} 
\\
\multicolumn{5}{l}{\textbf{11. Distressed}}
\\
A. Very Rarely or Not at All 
& B. Very Little 
& C. Moderately 
& D. Quite a Bit 
& E. Very Much
\\ [5pt]
\multicolumn{5}{l}{\textbf{12. Upset}}
\\ [5pt]
A. Very Rarely or Not at All 
& B. Very Little 
& C. Moderately 
& D. Quite a Bit 
& E. Very Much
\\ [5pt]
\multicolumn{5}{l}{\textbf{13. Guilty}}
\\
A. Very Rarely or Not at All 
& B. Very Little 
& C. Moderately 
& D. Quite a Bit 
& E. Very Much
\\ [5pt]
\multicolumn{5}{l}{\textbf{14. Scared}}
\\
A. Very Rarely or Not at All 
& B. Very Little 
& C. Moderately 
& D. Quite a Bit 
& E. Very Much
\\ [5pt]
\multicolumn{5}{l}{\textbf{15. Hostile}}
\\
A. Very Rarely or Not at All 
& B. Very Little 
& C. Moderately 
& D. Quite a Bit 
& E. Very Much
\\ [5pt] 
\multicolumn{5}{l}{\textbf{16. Irritable}}
\\
A. Very Rarely or Not at All 
& B. Very Little 
& C. Moderately 
& D. Quite a Bit 
& E. Very Much
\\ [5pt]
\multicolumn{5}{l}{\textbf{17. Ashamed}}
\\
A. Very Rarely or Not at All 
& B. Very Little 
& C. Moderately 
& D. Quite a Bit 
& E. Very Much
\\ [5pt]
\multicolumn{5}{l}{\textbf{18. Nervous}}
\\
A. Very Rarely or Not at All 
& B. Very Little 
& C. Moderately 
& D. Quite a Bit 
& E. Very Much
\\ [5pt]
\multicolumn{5}{l}{\textbf{19. Jittery}}
\\
A. Very Rarely or Not at All 
& B. Very Little 
& C. Moderately 
& D. Quite a Bit 
& E. Very Much
\\ [5pt]
\multicolumn{5}{l}{\textbf{20. Afraid}}
\\
A. Very Rarely or Not at All 
& B. Very Little 
& C. Moderately 
& D. Quite a Bit 
& E. Very Much
\\
\hline
\end{tabular}
\caption{The Questionnaire Measuring the Emotions of a Client based on PANAS.}
\label{table:questionnaire}
\end{table*}

\subsubsection{Testing Procedure}
\label{append:guidance}
The guidance is a clear, step-by-step guide, ensuring the client understands the purpose and process of the PANAS, and offering support throughout. This approach helps the client feel comfortable and understood, encouraging honest and accurate responses.

\noindent
\textbf{Introduction and Explanation}
(1) Introduce the Tool: "Today, I'd like to introduce you to a tool called the Positive and Negative Affect Schedule, or PANAS. It's a widely used measure in psychology to assess different aspects of your mood and emotions."
(2) Purpose: "The purpose of PANAS is to help us understand how you experience positive and negative feelings in your daily life. This can give us valuable insights into your emotional well-being."

\noindent
\textbf{Description and Instructions}
(3) Describe the Format: "PANAS consists of a list of words that describe different feelings and emotions. You will see words like \textit{interested}, \textit{distressed}, \textit{excited}, and so on."
(4) Time Frame: "I would like you to think about how you've felt over the past week and rate each emotion based on this. If you have not experienced a certain emotion at all, that is perfectly okay; just rate it accordingly."
(5) Demonstrate Rating: "Each emotion should be rated on a scale from 1 to 5, where 1 means \textit{very slightly or not at all}, and 5 means \textit{extremely}. For example, if you've felt \textit{alert} quite strongly this week, you might rate it a 4 or 5."

\noindent
\textbf{Completing the Schedule}
(6) Encourage a Relaxed Setting: "Please take your time to go through this and try to find a quiet moment where you can reflect on your feelings without interruptions."
(7) Emphasize Honesty and Spontaneity: "Your responses are completely confidential. It is important to be as honest and spontaneous as possible. There are no right or wrong answers here."

\noindent
\textbf{Support and Availability}
(8) Offer Support: "If you have any questions while you are filling this out, or if any of the emotions or ratings are not clear, please feel free to ask me."

\subsection{Clients PANAS Score changes}
\label{append:client_change}
In this chapter, we present a detailed examination of the clients' PANAS (Positive and Negative Affect Schedule) scores in Table \ref{table:client_change}, both before and after undergoing psychological therapy. This quantitative analysis aims to showcase the impact of the therapy on their emotional well-being. The corresponding radar chart, which visually represents these changes comprehensively, can be found in Table \ref{table:human_result}. This table not only illustrates the shifts in positive and negative affect but also provides a nuanced insight into the effectiveness of the therapeutic interventions applied.

\begin{table*}[hb]
\tiny
\centering
\renewcommand{\arraystretch}{1.3}
\scalebox{0.9}{
\begin{tabular}{p{7mm}p{1mm}p{1mm}p{2mm}p{0.5mm}p{1mm}p{1mm}p{2mm}p{0.5mm}p{1mm}p{1mm}p{2mm}p{0.5mm}p{1mm}p{1mm}p{2mm}p{0.5mm}p{1mm}p{1mm}p{2mm}p{0.5mm}p{1mm}p{1mm}p{2mm}p{0.5mm}p{1mm}p{1mm}p{2mm}p{0.5mm}p{1mm}p{1mm}p{2mm}}
\hline
& \multicolumn{7}{c}{ChatGLM3-6b} 
&
& \multicolumn{7}{c}{LLaMA2-7b-chat} 
&
& \multicolumn{7}{c}{HealMe} 
&
& \multicolumn{7}{c}{Control Group} 
\\ \cline{2-8} \cline{10-16} \cline{18-24} \cline{26-32}
& \multicolumn{3}{c}{Client 1} 
&  
& \multicolumn{3}{c}{Client 2} 
&  
& \multicolumn{3}{c}{Client 3} 
&  
& \multicolumn{3}{c}{Client 4} 
&  
& \multicolumn{3}{c}{Client 5} 
&  
& \multicolumn{3}{c}{Client 6}
&  
& \multicolumn{3}{c}{Client 7}
&  
& \multicolumn{3}{c}{Client 8}
\\ \cline{2-4} \cline{6-8} \cline{10-12} \cline{14-16} \cline{18-20} \cline{22-24} \cline{26-28} \cline{30-32}
& b        
& a    
& $\delta$
&  
& b        
& a  
& $\delta$
&  
& b    
& a 
& $\delta$
& 
& b
& a   
& $\delta$
&
& b  
& a     
& $\delta$
& 
& b      
& a     
& $\delta$
&
& b  
& a     
& $\delta$
& 
& b      
& a     
& $\delta$
\\ 
\cline{2-4}
\cline{6-8}
\cline{10-12}
\cline{14-16} 
\cline{18-20}
\cline{22-24} 
\cline{26-28}
\cline{30-32}
Interested  
& 3         
& 4     
& +1
&  & 2       
& 3   
& +1
&  & 2       
& 2   
& $\backslash$
&  & 3       
& 2    
& -1
&  & 2       
& 3         
& +1
&  & 3        
& 3     
& $\backslash$
& & 2
& 2
& $\backslash$
& & 4
& 4
& $\backslash$
\\
Excited   
& 3      
& 3  
& $\backslash$
&  & 2       
& 3    
& +1
&  & 2       
& 1    
& -1
&  & 2      
& 2    
& $\backslash$
&  & 2     
& 2    
& $\backslash$
&  & 3     
& 3  
& $\backslash$
& & 2
& 2
& $\backslash$
& & 4
& 4
& $\backslash$
\\
Strong   
& 3     
& 4    
& +1
&  & 2    
& 4   
& +2
&  & 2     
& 1   
& -1
&  & 2     
& 2     
& $\backslash$
&  & 3     
& 3    
& $\backslash$
&  & 2     
& 4    
& +2
& & 3
& 3
& $\backslash$
& & 2
& 2
& $\backslash$
\\
Enthusiastic
& 3       
& 4    
& +1
&  & 2    
& 4  
& +2
&  & 1    
& 1    
& $\backslash$
&  & 2   
& 2  
& $\backslash$
&  & 3   
& 3     
& $\backslash$
&  & 3   
& 3  
& $\backslash$
& & 2
& 3
& +1
& & 1
& 1
& $\backslash$
\\
Proud   
& 3      
& 3    
& $\backslash$
&  & 3    
& 4   
& +1
&  & 1   
& 1 
& $\backslash$
&  & 2    
& 2
& $\backslash$
&  & 3    
& 2     
& -1
&  & 3    
& 3       
& $\backslash$
& & 3
& 3
& $\backslash$
& & 2
& 2
& $\backslash$
\\
Alert   
& 4     
& 2     
& -2
&  & 3   
& 2     
& -1
&  & 1   
& 1   
& $\backslash$
&  & 3    
& 3    
& $\backslash$
&  & 4    
& 3     
& -1
&  & 5     
& 3       
& -2
& & 2
& 2
& $\backslash$
& & 2
& 2
& $\backslash$
\\
Inspired 
& 3        
& 3       
& $\backslash$
&  & 2      
& 3        
& +1
&  & 1     
& 1       
& $\backslash$
&  & 2     
& 2    
& $\backslash$
&  & 3     
& 3     
& $\backslash$
&  & 2     
& 3    
& +1
& & 3
& 3
& $\backslash$
& & 1
& 1
& $\backslash$
\\
Determined 
& 3      
& 4     
& +1
&  & 3   
& 3      
& $\backslash$
&  & 1    
& 1    
& $\backslash$
&  & 2   
& 2   
& $\backslash$
&  & 2    
& 3       
& +1
&  & 3   
& 4   
& +1
& & 3
& 3
& $\backslash$
& & 4
& 3
& -1
\\
Attentive 
& 4       
& 4    
& $\backslash$
&  & 3   
& 3      
& $\backslash$
&  & 1   
& 1      
& $\backslash$
&  & 1    
& 1       
& $\backslash$
&  & 2     
& 2       
& $\backslash$
&  & 5    
& 3    
& -2
& & 1
& 2
& +1
& & 2
& 2
& $\backslash$
\\
Active   
& 3      
& 4  
& +1
&  & 2   
& 3   
& +1
&  & 1   
& 1     
& $\backslash$
&  & 2    
& 2    
& $\backslash$
&  & 3    
& 3    
& $\backslash$
&  & 4    
& 3       
& -1
& & 2
& 2
& $\backslash$
& & 1
& 1
& $\backslash$
\\ \hline
Distressed 
& 3       
& 2       
& -1
&  & 4    
& 1       
& -3
&  & 4    
& 4       
& $\backslash$
&  & 3     
& 4        
& +1
&  & 4    
& 4      
& $\backslash$
&  & 5     
& 1     
& -4
& & 3
& 3
& $\backslash$
& & 5
& 5
& $\backslash$
\\
Upset    
& 3      
& 2     
& -1
&  & 2    
& 1      
& -1
&  & 4    
& 5     
& +1
&  & 4     
& 3   
& -1
&  & 4    
& 3      
& -1
&  & 5     
& 1      
& -4
& & 4
& 3
& -1
& & 5
& 5
& $\backslash$
\\
Guilty    
& 3      
& 1      
& -2
&  & 1  
& 1    
& $\backslash$
&  & 2    
& 1    
& -1
&  & 1    
& 1   
& $\backslash$
&  & 3        
& 2      
& -1
&  & 1     
& 1         
& $\backslash$
& & 1
& 1
& $\backslash$
& & 1
& 2
& +1
\\
Scared  
& 3      
& 1      
& -2
&  & 1   
& 1  
& $\backslash$
&  & 4   
& 2      
& -2
&  & 3   
& 2    
& -1
&  & 1    
& 1   
& $\backslash$
&  & 1     
& 1 
& $\backslash$
& & 2
& 2
& $\backslash$
& & 4
& 4
& $\backslash$
\\
Hostile  
& 2       
& 1    
& -1
&  & 1   
& 1    
& $\backslash$
&  & 4    
& 5     
& +1
&  & 1    
& 2     
& +1
&  & 1    
& 1       
& $\backslash$
&  & 5    
& 3   
& -2
& & 1
& 1
& $\backslash$
& & 5
& 5
& $\backslash$
\\
Irritable   
& 3         
& 2        
& -1
&  & 3     
& 2        
& -1
&  & 5      
& 5       
& $\backslash$
&  & 3    
& 4        
& +1
&  & 3      
& 4  
& +1
&  & 5     
& 1    
& -4
& & 4
& 3
& -1
& & 5
& 5
& $\backslash$
\\
Ashamed   
& 2      
& 1    
& -1
&  & 1   
& 1    
& $\backslash$
&  & 3     
& 2       
& -1
&  & 1      
& 1     
& $\backslash$
&  & 1    
& 1  
& $\backslash$
&  & 1     
& 1   
& $\backslash$
& & 2
& 1
& -1
& & 2
& 1
& -1
\\
Nervous   
& 3      
& 1     
& -2
&  & 3  
& 2     
& -1
&  & 3  
& 4   
& +1
&  & 3   
& 3   
& $\backslash$
&  & 3  
& 4   
& +1
&  & 2  
& 1    
& -1
& & 3
& 3
& $\backslash$
& & 3
& 3
& $\backslash$
\\
Jittery  
& 3       
& 2     
& -1
&  & 4  
& 2     
& -2
&  & 5  
& 5      
& $\backslash$
&  & 4   
& 4 
& $\backslash$
&  & 4  
& 4 
& $\backslash$
&  & 4    
& 1   
& -3
& & 4
& 4
& $\backslash$
& & 5
& 5
& $\backslash$
\\
Afraid   
& 3      
& 1     
& -2
&  & 1    
& 1      
& $\backslash$
&  & 4     
& 2    
& -2
&  & 3       
& 3     
& $\backslash$
&  & 3     
& 4     
& +1
&  & 4      
& 1    
& -3
& & 3
& 4
& +1
& & 3
& 3
& $\backslash$
\\ \hline
\end{tabular}}
\caption{Changes in PANAS Scores for Eight Clients Pre- and Post-Intervention. Notation: \textit{b} indicates scores before the intervention, \textit{a} represents scores after the intervention, and $\delta$ denotes the change calculated as post-intervention scores minus pre-intervention scores.}
\label{table:client_change}
\end{table*}

Based on statistics in Table \ref{table:human_result}, we analyze the effectiveness of the PANAS test and make a comparison of reduction in negative emotions. 

The average fluctuation in negative emotions (absolute value of change in negative emotions/total negative emotion score before the experiment) for the control and three experimental groups pre- and post-test are as follows:
\begin{itemize}
    \item ChatGLM Group: 44\%
    \item LLaMA Group: 21\%
    \item HealMe Group: 41\%
    \item Control Group: 10\%
\end{itemize}

This finding demonstrates a strong correlation between the use of the models and the fluctuation in clients' negative emotions, observed before and after the dialogue sessions. 

We calculate the standard deviation of the negative emotion change scores for the four groups:
\begin{itemize}
    \item ChatGLM Group: 0.77
    \item LLaMA Group: 0.95
    \item HealMe Group: 1.23
    \item Control Group: 0.55
\end{itemize}

This observation shows our model's effectiveness in alleviating clients' negative emotions. This finding underscores the potential of our model in psychotherapy.

\subsection{Client's feedback on baseline models}
\label{sec:feedback}
Clients 3 and 4 mention that the model (LLaMA2-7b-chat) offers cold advice during conversations, with Client 3 wishing for more empathetic support. While following the advice of this model, Client 3 also experiences reduced feelings of fear and can face emotions more objectively. Clients 1 and 2, who have less negative emotion, seek advice from the model (ChatGLM3-6b) and are satisfied with the suggestions, leading to a decrease in negative emotions. However, Client 1 notes that in his case, the model (ChatGLM3-6b) lacks strong guidance and feels more like a search engine than a psychological therapist.

\end{document}